\documentclass[aps,reprint,amsmath,amssymb,groupaddress,superscriptaddress,prl]{revtex4-2}

\usepackage[utf8]{inputenc}
\usepackage{amsmath}
\usepackage{amsfonts}
\usepackage{amssymb}
\usepackage{bm}
\usepackage{color}
\usepackage{graphicx}
\usepackage{soul}
\usepackage{hyperref}
\usepackage{CJK}
\usepackage{xcolor}
\usepackage{mathtools}
\usepackage{lipsum}

\begin{document}
\begin{CJK*}{UTF8}{gbsn}

\title{A Minimal Model for Carnot Efficiency at Maximum Power}
\author{Shiling Liang (梁师翎)}
 \email{shiling.liang@epfl.ch}
\affiliation{Institute of Physics, School of Basic Sciences, \'Ecole Polytechnique F\'ed\'erale de Lausanne (EPFL), 1015 Lausanne, Switzerland}
\affiliation{Biological Complexity Unit, Okinawa Institute of Science and Technology Graduate University, Onna, Okinawa 904-0495, Japan}
\author{Yu-Han Ma}
 \email{yhma@bnu.edu.cn}
\affiliation{School of Physics and Astronomy, Beijing Normal University, Beijing, 100875, China}
\affiliation{Graduate School of China Academy of Engineering Physics, No. 10 Xibeiwang East Road, Haidian District, Beijing, 100193, China}
\author{Daniel Maria Busiello}
\affiliation{Max Planck Institute for the Physics of Complex Systems, 01187 Dresden, Germany}
\author{Paolo De Los Rios}
\affiliation{Institute of Physics, School of Basic Sciences, \'Ecole Polytechnique F\'ed\'erale de Lausanne (EPFL), 1015 Lausanne, Switzerland}
\affiliation{Institute of Bioengineering, School of Life Sciences, \'Ecole Polytechnique F\'ed\'erale de Lausanne (EPFL), 1015 Lausanne, Switzerland}

\begin{abstract}
Carnot efficiency sets a fundamental upper bound on the heat engine efficiency, attainable in the quasi-static limit, albeit at the cost of completely sacrificing power output. In this Letter, we present a minimal heat engine model that can attain Carnot efficiency while achieving maximum power output. We unveil the potential of intrinsic divergent physical quantities within the working substance, such as degeneracy, as promising thermodynamic resources to break through the universal power-efficiency trade-off imposed by nonequilibrium thermodynamics for conventional heat engines. Our findings provide novel insights into the collective advantage in harnessing energy of many-body interacting systems.
\end{abstract}

\maketitle
\end{CJK*}
Heat engines generally operate between two thermal baths at different temperatures ($T_h>T_c$), converting thermal energy into output work. Their efficiency, $\eta = W/Q_h$, i.e., the ratio of output work $W$ to absorbed heat $Q_h$, is bounded by the second law of thermodynamics, with an upper limit defined by the Carnot efficiency $\eta_C = 1-T_c/T_h$ \cite{Callen1991thermodynamics}. Since any finite-time operation in a thermodynamic system leads to irreversible entropy production, the Carnot limit can only be achieved when an engine operates in a quasi-static regime, resulting in zero power output \cite{shiraishi2018Speeda, salamon1983Thermodynamic}. However, functional heat engines require both high efficiency and high output power. To evaluate their performance, efficiency at maximum power (EMP) \cite{NoVIKOV1958,curzon1975Efficiency, vandenbroeck2005Thermodynamic,Schmiedl2008,tu2008Efficiency,esposito2010Efficiency} and the trade-off relation between power and efficiency \cite{chen1989effect,holubec2016Maximum,ma2018Universal,yuan2022Optimizing,zhai2023experimental} have been proposed as key benchmarks, drawing significant attention in nonequilibrium and engineering thermodynamics \cite{andresen2011current,tu2021abstract,berry2022FiniteTime}.

Recent developments in stochastic thermodynamics paved new ways to investigate heat engines operating far from equilibrium \cite{shiraishi2023Introduction}. These developments offer the potential to achieve Carnot efficiency with finite \cite{polettini2017Carnot,lee2017Carnot,abiuso2020Optimal,campisi2016Power} or maximum  \cite{seifert2011Efficiency,allahverdyan2013Carnot} power using working substances with intrinsic divergent physical quantities. Many-body interacting systems exploit these properties through emerging collective effects, significantly reducing the irreversible dissipation relative to output work and thus enhancing engine efficiency \cite{campisi2016Power,rolandi2023Collective,vroylandt2018Collective,filho2023Powerful,tajima2021Super}. Despite significant progress, a comprehensive understanding of the key advantages stemming from collective operations remains elusive. Key open questions include the existence of a universal power-efficiency trade-off for collective heat engines, and the relationship between engine performance and many-body phase transitions.

\begin{figure}[!htb]
    \includegraphics[width=1\columnwidth]{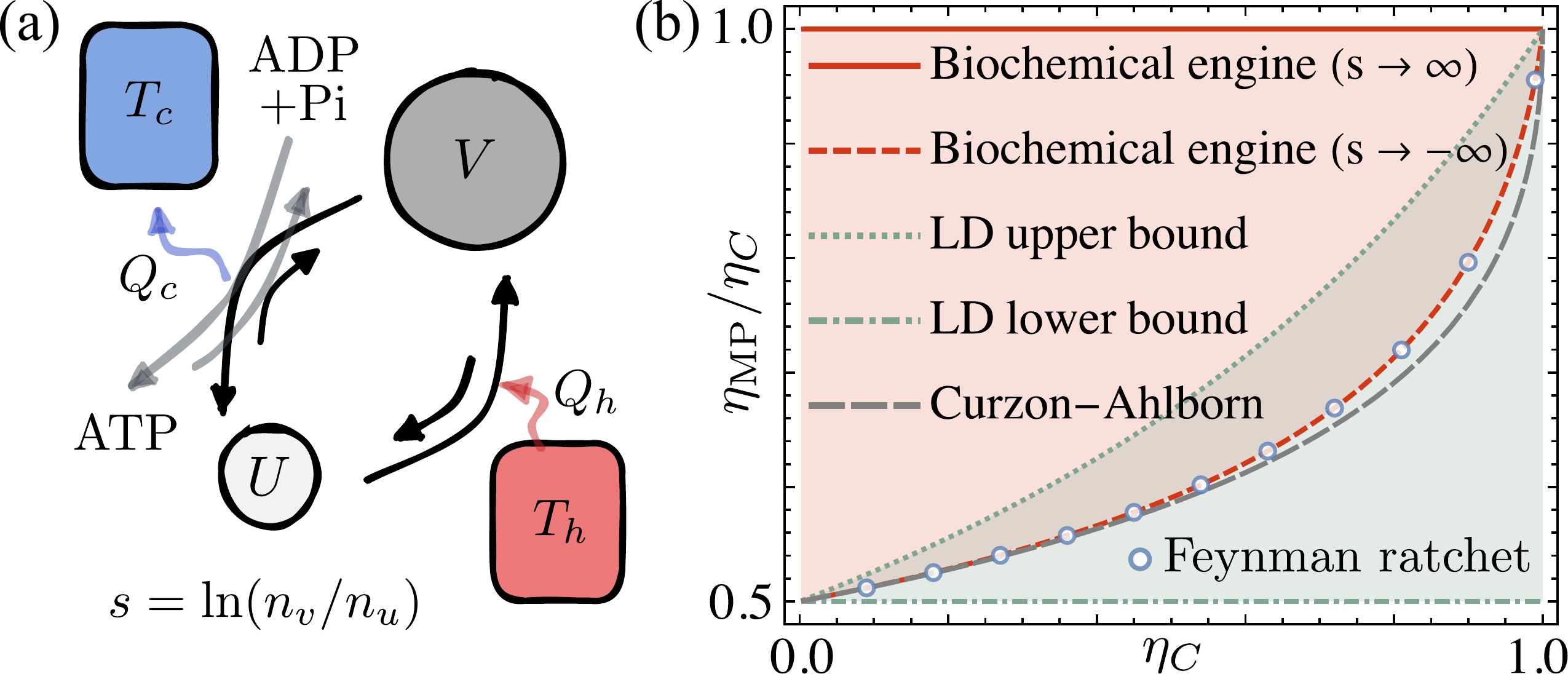}
    \caption{(a) A biochemical engine operating between two thermal baths to convert thermal into chemical energy through ATP synthesis. $U$ and $V$ are two coarse-grained states with degeneracies $n_u$ and $n_v$, resulting in an entropy difference $s=\ln (n_v/n_u)$. (b) The EMP of this engine can surpass the universal upper bound for low-dissipation (LD) engines \cite{esposito2010Efficiency} and reach Carnot efficiency at $s\to\infty$.}
    \label{fig: heat_engine}
\end{figure}

\begin{figure*}[!htb]    \includegraphics[width=2\columnwidth]{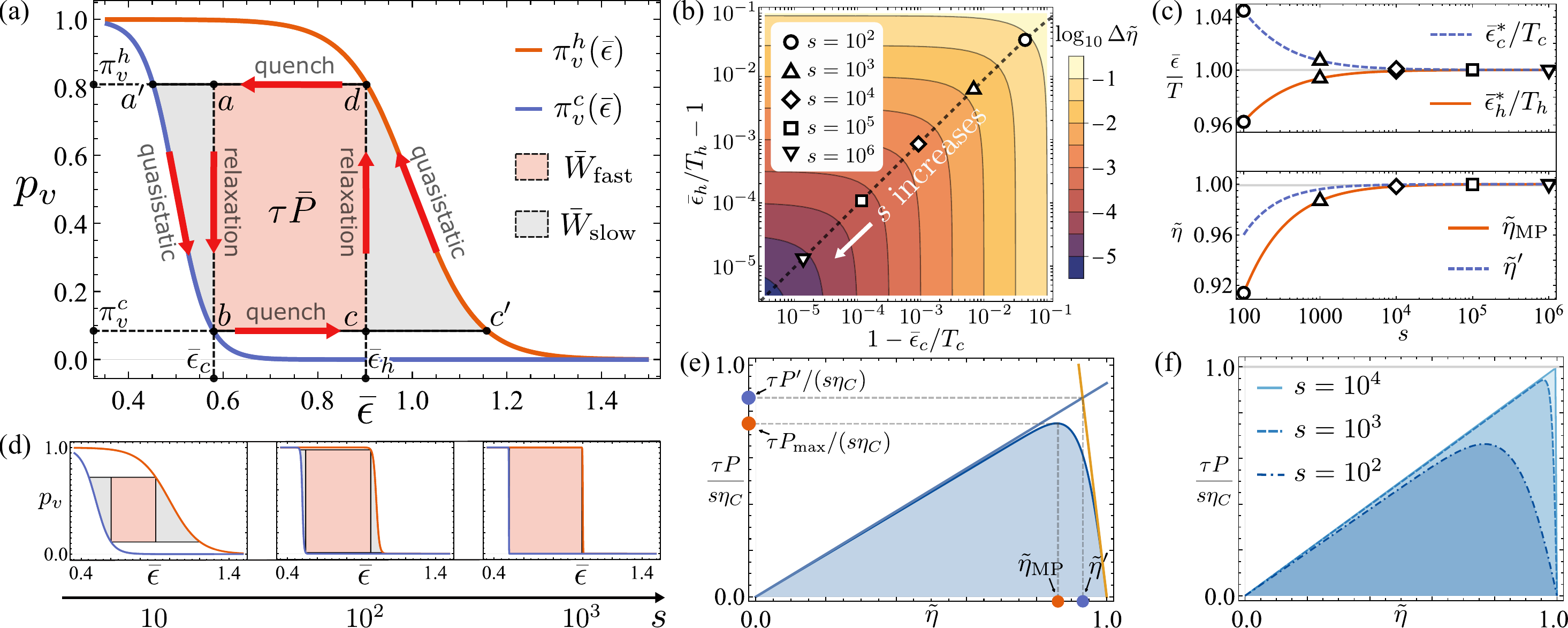}
    \caption{Optimization of the biochemical engine. (a) An Otto cycle, $abcda$, operates between two thermal reservoirs. Red and blue curves denote the equilibrium population on state $v$ coupled to $T_h$ and $T_c$, respectively. $a'bc'da'$ is a Carnot cycle. (b) Deviation from Carnot efficiency $\Delta\tilde{\eta}=1-\eta/\eta_C$ as a function of thermodynamic parameters $\bar{\epsilon}_h/T_h$ and $\bar{\epsilon}_c/T_c$. Points and dashed line indicate optimal parameters, $\bar{\epsilon}_h^*$ and $\bar{\epsilon}_c^*$, that maximize the power for different $s$. (c) In the large-$s$ limit, maximum power is reached when $\bar{\epsilon}_{c/h}^*/T_{c/h}\to 1$, while the rescaled EMP approaches the Carnot limit, $\tilde{\eta}_\mathrm{MP}\equiv\eta_\mathrm{MP}/\eta_C \to 1$. (d) Optimized Otto cycles to attain maximum output work under different values of $s$. (e) Power-efficiency trade-off and linear expansion lines from the two endpoints $\tilde{\eta}\to 0$ and $\tilde{\eta}\to 1$. The intersection of these lines gives upper bounds on maximum power, $P'$, and corresponding efficiency, $\eta'$. (f) Power-efficiency trade-off at different $s$, obtained via numerical optimization \cite{optimization}, reveals power scaling with $s\eta_C$ at large s. We set $k_BT_h=1$ as the reference energy and $\eta_C=0.5$.}
\label{fig:optimization}
\end{figure*}

\begin{figure}[!htb]
\includegraphics[width=1\columnwidth]{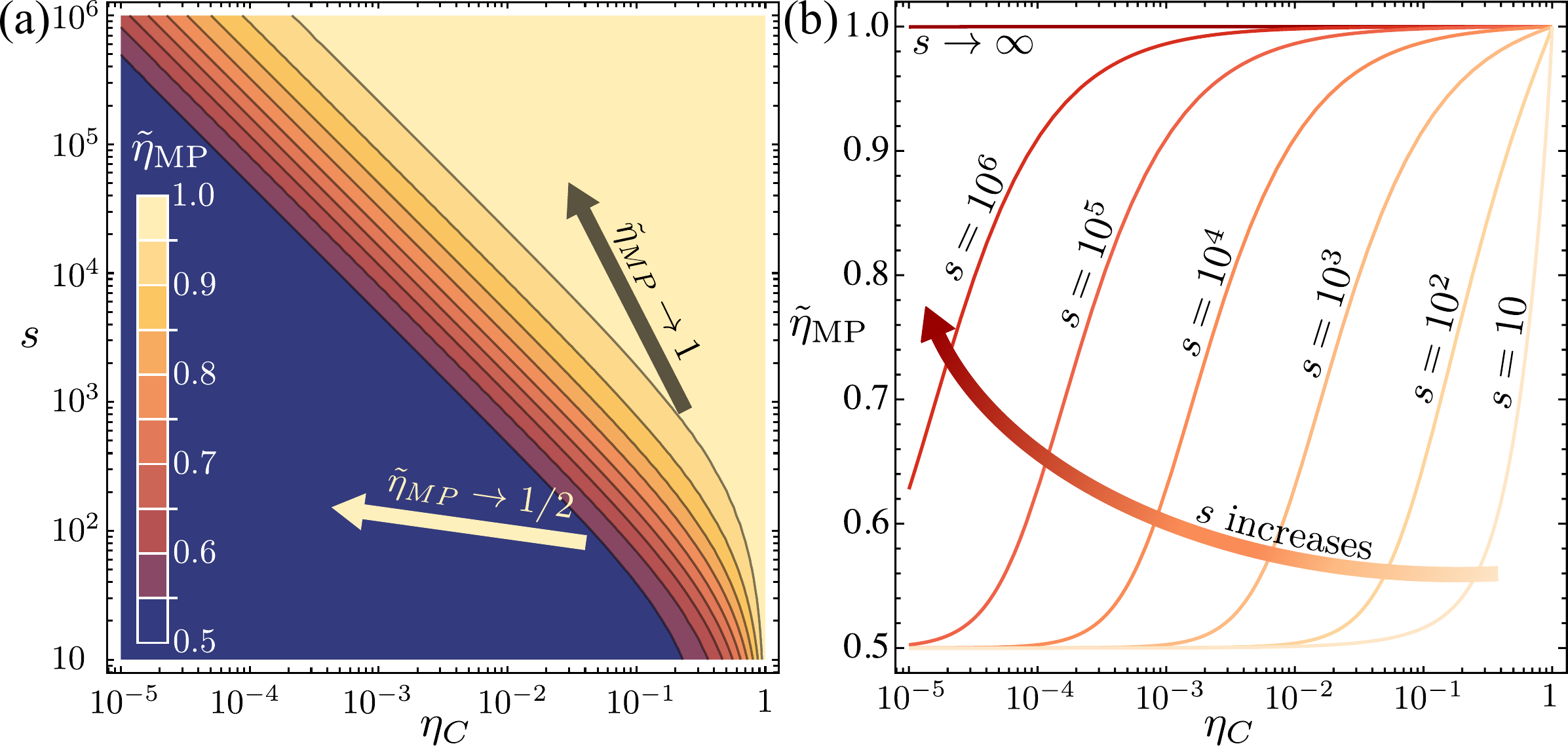}
\caption{(a) Rescaled EMP, $\tilde{\eta}_{\rm MP}\equiv \eta_{\rm MP}/\eta_C$, approaches $1$ when $s\rightarrow\infty$ faster than $\eta_C\rightarrow0$, breaking the $1/2$-universality. (b) $\tilde{\eta}_\mathrm{MP}$ as a function of $\eta_C$ for different $s$.}
    \label{fig:1/2-universality}
\end{figure}

To address these problems, we present a minimal model  (Fig.~\ref{fig: heat_engine}a) that incorporates intrinsic energy level degeneracy, thereby circumventing the complexity while capturing the basic degenerate characteristics arising from many-body interactions. This particular engine can achieve Carnot efficiency at maximum power (Fig.~\ref{fig: heat_engine}b). More importantly, we analytically derive the asymptotic behavior of the power-efficiency trade-off (Fig.~\ref{fig:optimization}f) in the limit of large degeneracy of the high-energy state. Our findings explicitly demonstrate the violation of the $1/2$-universality of EMP \cite{vandenbroeck2005Thermodynamic,Tu2012Recent} in the thermodynamic limit (Fig.~\ref{fig:1/2-universality}). We show that the thermodynamic advantage of the proposed engine stems from the first-order phase transition occurring far from the linear response regime.

\textit{Model}.---
Mesoscopic heat engines can produce work by synthesizing ATP \cite{seifert2011Efficiency,grelier2024Unlocking}. In this Letter, we consider a biochemical engine composed of a low-energy state $U$ and a high-energy state $V$ representing two coarse-grained states with $n_u$ and $n_v$ microscopic conformations, respectively (Fig.~\ref{fig: heat_engine}a). For example, $U$ and $V$ can represent the folded and unfolded states of a polymer \cite{maffi2012Firstorder,liang2022emergent} or chemical species \cite{dass2021equilibrium}. The entropy difference between these two coarse-grained states, $s=\ln(n_v/n_u)$, comes from the degeneracy difference. There are two reaction paths between $U$ and $V$: an ATP-hydrolysis-driven reaction under low temperature $T_c$, and a spontaneous transition coupled to the high-temperature reservoir $T_h$,
\begin{displaymath}
    \begin{aligned}
    U \xrightleftharpoons[k_{h-}]{k_{h+}}
     V,\quad
     \mathrm{ATP} + U
     \xrightleftharpoons[k_{c-}^0]{k_{c+}^0}
     V+\mathrm{ADP} + \mathrm{Pi}.
     \end{aligned}
\end{displaymath}
Here $\{k_{c+}^0,k_{c-}^0\}$ denote the forward and backward ATP-driven reaction rates, and $\{k_{h+},k_{h-}\}$ are the high-temperature spontaneous transition rates. Absorbing the chemostatted concentrations of ATP, ADP and Pi into the rates, we can define effective rates $k_{c+} = k_{c+}^0[\mathrm{ATP}]$ and $k_{c-}=k_{c-}^0[\mathrm{ADP}][\mathrm{Pi}]$. All rates must satisfy local detailed balance, $k_{h-}/k_{h+}=e^{\beta_h \epsilon_h-s}$ and $k_{c-}/k_{c+}=e^{\beta_c \epsilon_c-s}$, where $\beta_{h/c} = 1/k_B T_{h/c}$ \cite{maes2021Local,peliti2021stochastic}. Here, $\epsilon_h$ and $\epsilon_c$ are the energy gaps of the system when coupled to the hot and cold reservoirs, respectively. The change of energy gap $\Delta \mu \equiv \epsilon_h-\epsilon_c$ 
is the output work per cycle in terms of ATP synthesis. 

The dynamics of the system is described by the master equation
\begin{equation}\label{eq:master}
    \begin{aligned}
       \frac{d}{dt}p_{u} &= (k_{h-} + k_{c-})p_{v}- (k_{h+}+ k_{c+})p_u ,\\
       \frac{d}{dt}p_{v} &=  (k_{h+}+ k_{c+})p_u-(k_{h-} + k_{c-})p_{v}.
    \end{aligned}
\end{equation}
where $p_u$ and $p_v$ are the probabilities of two states. Each reaction path has a corresponding equilibrium distribution, 
\begin{equation}\label{eq:eq_dis}
\pi_v^{h/c}=\frac{1}{1+e^{\beta_{h/c} \epsilon_{h/c}-s}},\quad \pi_u^{h/c}=\frac{e^{\beta_{h/c} \epsilon_{h/c}-s}}{1+e^{\beta_{h/c} \epsilon_{h/c}-s}}.
\end{equation}
that satisfy $\pi_u^h/\pi_v^h=k_{h-}/k_{h+}$ and $\pi_u^c/\pi_v^c=k_{c-}/k_{c+}$. 

At stationarity,  $p_{u/v}^\mathrm{ss} = (\tau_h \pi_{u/v}^{c} +\tau_c\pi_{u/v}^{h})/(\tau_h+\tau_c)$ as a consequence of the competition between two reaction paths with typical time-scales $\tau_h=(k_{h+}+k_{h-})^{-1}$ and $\tau_c=(k_{c+}+k_{c-})^{-1}$. The corresponding steady-state cyclic current is \cite{supplemental_material}
\begin{equation}\label{eq:current}
    J^\mathrm{ss}  =\tau^{-1} (\pi_{v}^{h} - \pi_{v}^{c}),
\end{equation}
where $\tau = \tau_h+\tau_c$ is the characteristic time of completing the cycle.
The change of internal entropy does not lead to entropy production into the environment \cite{peliti2021stochastic,esposito2012Stochastic,seifert2011Stochastic}, and thus the heat absorption rate from the hot (cold) reservoir is given by $\dot{Q}_{h}= J^\mathrm{ss}\epsilon_{h}$ ($ \dot{Q}_{c}= -J^\mathrm{ss}\epsilon_{c}$). The steady-state output power is
\begin{equation}\label{eq:power_def}
       {P} = J^\mathrm{ss}(\epsilon_h-\epsilon_c) = \tau^{-1}(\pi_v^h-\pi_v^c)(\epsilon_h-\epsilon_c),
\end{equation}
then the thermodynamic efficiency reads
\begin{equation}\label{eq:efficiency}
    \eta=P/\dot{Q}_h=1-(\epsilon_c/\epsilon_h).
\end{equation}
One can find that the quasi-static limit, i.e. $J^\mathrm{ss}=0$, is reached when $\epsilon_c T_h=\epsilon_h T_c$ and leads to Carnot efficiency.

\textit{Optimization}.---
To maximize the power of the biochemical engine, we set the entropy difference $s$ as a control parameter and optimize the two energy levels $\epsilon_c$ and $\epsilon_h$ under different $T_h$ and $T_c$. For convenience, we set the Boltzmann constant $k_B=1$, and use $k_BT_h=1$ as the reference energy. We follow the convention to assume the characteristic time $\tau$ constant because it is a kinetic parameter and thus not constrained by thermodynamics in general cases \cite{tu2008Efficiency}. Therefore, we optimize the work of the thermodynamic cycle,
\begin{equation}
    W = \tau P=(\pi_v^h-\pi_v^c)(\epsilon_h-\epsilon_c) \equiv W_{\mathrm{Otto}},
\end{equation}
which has an intuitive geometric representation in the state space, as shown in Fig.~\ref{fig:optimization}a. Indeed, it corresponds to the work of an Otto cycle $abcda$ operating between two thermal reservoirs (see \cite{supplemental_material} for a detailed comparison). Hereafter, we rescale physical quantities by $s$ such that $\bar{x}\equiv x/s$, since $s$ characterizes the system size, as discussed below. In the Otto cycle, the two quench processes $d\to a$ and $b\to c$ do not have intrinsic speed limits while the two isothermal relaxations $c\to d$ and $a\to b$ have intrinsic relaxation timescales $\tau_{{h/c}}=k_{h/c}^{-1}$ that set $\tau = \tau_h + \tau_c$ as the time of completing the cycle \cite{supplemental_material}. 

Let us start with limiting cases. In Fig.~\ref{fig: heat_engine}b, we present EMP, $\eta_{\rm MP}$, as a function of $\eta_C$. When $s\to -\infty$, we can expand $p_{u}^\mathrm{ss}$ with respect to $e^s$, so that the power becomes ${P} =[T_c e^s\left(e^{-\theta(1-\eta)}  -e^{-\theta (1-\eta_C)}\right)\eta\theta]/\tau$, where $ \theta\equiv \epsilon_h\beta_c$. This expression has the same mathematical structure as Feynman's ratchet heat engine whose closed-form EMP has been obtained in \cite{tu2008Efficiency}. The resulting EMP (blue dots in Fig.~\ref{fig: heat_engine}b) serves as the lower bound for the biochemical engine and is higher than the Curzon-Ahlborn efficiency \cite{curzon1975Efficiency}.

For large $s$, i.e., high degeneracy of the high-energy state, a closed-form EMP cannot be obtained. However, numerical optimization of Eq.~\eqref{eq:power_def} \cite{optimization} shows that $\eta_{\rm MP}$ surpasses the universal upper bound $\eta_C/(2-\eta_C)$ of low-dissipation heat engines \cite{esposito2010Efficiency} and approaches the Carnot efficiency as $s\to \infty$ (Fig.\ref{fig: heat_engine}b, red solid line). In Fig.~\ref{fig:optimization}b, we illustrate the efficiency of the biochemical engine as a function of the thermodynamic parameters $\bar{\epsilon}_h/T_h$ and $\bar{\epsilon}_c/T_c$. The color map represents the relative deviation of efficiency from $\eta_C$, $\Delta\tilde{\eta}\equiv1-\eta/\eta_C$, which approaches $0$ in the $s\to \infty$ limit. Markers and dashed line indicate $(\bar{\epsilon}_h^*,\bar{\epsilon}_c^*)$ that maximize the power at different $s$. Figs.~\ref{fig:optimization}b-c together indicate that the optimizations of power and efficiency coincide in the large-$s$ limit and lead to Carnot efficiency at maximum power.

\textit{Thermodynamic advantage of phase transition}.--- 
As illustrated in Fig.~\ref{fig:optimization}c, in the large-$s$ limit, the maximum power is achieved when $\bar{\epsilon}_{h/c}^*/T_{h/c}\to 1$. These optimized energy gaps correspond to the most sensitive region of the equilibrium curves $\pi_v^{h/c}(\bar{\epsilon})$, as shown in Fig.~\ref{fig:optimization}d. Since the slope,
\begin{equation}\label{eq:response}
\left. \frac{{\rm{d}}\pi_v^{h/c}(\bar{\epsilon})}{\rm{d}\bar{\epsilon}}\right|_{\bar{\epsilon}=T_{h/c}}=-\frac{s}{4T_{h/c}},
\end{equation}
scales linearly with $s$, when $s\to\infty$, $T_{h/c}=\bar{\epsilon}_{h/c}$ are two phase-transition points with diverging slope. At infinite $s$, when $\bar{\epsilon}_{h}$ crosses this point, the equilibrium distribution $\pi^{h}_{v}$ transitions from $0$ to $1$ (Fig.~\ref{fig:optimization}d, red line), and the same holds for $\pi_v^c$ (Fig.~\ref{fig:optimization}d, blue line), consequently inducing an abrupt change in $p_v$ during relaxation.

To understand the physical meaning of the $s\to \infty$ limit, we consider the simple example of $N$ interacting units that can be in an inactive ($\sigma_i = 0$) or active ($\sigma_i=1$) state and exhibit $N$-body interactions with the Hamiltonian  $H = -\epsilon \prod_{i=1}^N \sigma_i $. This system has two coarse-grained states: a low-energy one, $U$, with $E_U = -\epsilon$ where all units are active, and a high-energy state, $V$, with $E_V = 0$ with at least one inactive unit. Their degeneracies are $n_u = 1$ and $n_v = 2^N - 1$, respectively. Their entropy difference $s = \ln(n_v/n_u) \simeq N \ln 2$ for large $N$, thus scaling linearly with $N$ and diverging at the thermodynamic limit $N\to \infty$. By coupling this system to two reservoirs, we recover our minimal model.
While this $N$-body interaction is idealized, it captures the essential physics. Importantly, recent studies show that properly designed 2-body interacting systems can exhibit similar collective advantage~\cite{rolandi2023Collective,abiuso2024Optimal}, suggesting a broad applicability for our findings.
 
A geometrical analysis of thermodynamic cycles at increasing $s$ will clarify the role of the phase transition in enhancing engine performance.
In the Otto cycle, irreversible dissipation originates from two relaxation processes and is quantified by the Kullback-Leibler divergence between their corresponding initial and final states \cite{kawai2007Dissipation,parrondo2015Thermodynamics}. To prevent irreversible dissipation, the Carnot cycle $a'bc'da'$ should be employed. It includes two quasi-static processes whose associated work is denoted by $W_{\rm slow}$ (Fig.~\ref{fig:optimization}a, where $\bar{W}_{\rm slow/fast}$ represents the work rescaled by $s$). The quality of the Otto cycle can be assessed by defining the quality factor $\mathcal{Q}$ as the ratio of its work, $W_\mathrm{Otto} \equiv W_\mathrm{fast}$ to the work of the Carnot cycle, $W_\mathrm{Carnot} \equiv W_\mathrm{fast}+W_\mathrm{slow}$. When $s\gg 1$, we find \cite{supplemental_material}
\begin{equation}
\mathcal{Q}\equiv \frac{W_\mathrm{Otto}}{W_\mathrm{Carnot}}=\frac{W_\mathrm{fast}}{W_\mathrm{fast}+W_\mathrm{slow}}\simeq\frac{\bar{\epsilon}_h-\bar{\epsilon}_c}{T_h-T_c}\Bigg|_{s\gg 1} \;.
\end{equation}
Thus, the power-optimized biochemical engine operates close to the phase transition points in both reservoirs, i.e., $T_{h/c}/\bar{\epsilon}^*_{h/c}\to 1$, $\mathcal{Q}$ approaches 1, 
and the Otto cycle coincides with the Carnot cycle (Fig.~\ref{fig:optimization}d).

We conclude that the first-order phase transition induced by the divergence of high-energy level degeneracy is a favorable thermodynamic resource, allowing the EMP to approach Carnot efficiency. Previous studies reporting thermodynamic advantages stemming from phase transitions were limited to finite power \cite{campisi2016Power}, quasi-static cycles \cite{ma2017quantum}, and slow-driving regimes \cite{abiuso2020Optimal}. 
We emphasize that our model not only pinpoints the minimal ingredients that capture the thermodynamic advantage of phase transition but also suggests that this advantage persists even in fast-driving scenarios, where the protocol involves effective quenching processes which are infinitely faster than the intrinsic engine time-scale.


\textit{Power-efficiency trade-off}.---
Substituting Eqs.~\eqref{eq:eq_dis} and \eqref{eq:efficiency} into Eq.~\eqref{eq:power_def}, the power can be written in terms of $\epsilon_h$ and the rescaled efficiency $\tilde{\eta}\equiv \eta/\eta_C$ as 
\begin{equation}\label{eq:Pofeta}
P = \tilde{\eta}\epsilon_h\tau^{-1}\left(\frac{\eta_C}{e^{ \epsilon_h-s}+1}-\frac{\eta_C}{e^{{\frac{\epsilon_h\left(1-\tilde{\eta}\eta_C\right)}{1-\eta_C}-s}}+1}\right).
\end{equation}
The power-efficiency trade-off curve is obtained by numerically maximizing $P$ with respect to $\epsilon_h$ \cite{optimization}, i.e., $\partial_{\epsilon_h} P(\epsilon_h,\tilde{\eta};\eta_C,s) = 0$ \cite{shiraishi2016Universal,holubec2016Maximum,ma2018Universal,yuan2022Optimizing} (Figs.\ref{fig:optimization}e-f). By expanding around the ends $\tilde{\eta}\approx 0$ and $\tilde{\eta}\approx 1$ for large $s$ (Fig.~\ref{fig:optimization}e) \cite{supplemental_material}, we obtain the following limiting behaviors for the power-efficiency trade-off:
\begin{equation}\label{eq:pe_trade_expan}
\begin{cases}
    \lim_{\tilde{\eta}\to 0}P \simeq \tau^{-1}  (s -\ln s )\eta_C  \tilde{\eta}, \\ 
     \lim_{\tilde{\eta}\to 1}P\simeq \frac{\tau^{-1} s ^2 \eta_C^2}{4(1- \eta_C)} (1-\tilde{\eta}).
     \end{cases}
\end{equation}
As shown in Fig.~\ref{fig:optimization}f, when $s\to \infty$, the shape of the trade-off curve approaches an isosceles right triangle in the rescaled parametric space, showing that maximum power and Carnot efficiency are achieved simultaneously.

To elucidate this behavior, we approximate the maximum power and EMP with the intersection points of the two expansions in Eq.~\eqref{eq:pe_trade_expan}, named $P'$ and $\tilde{\eta}'$ respectively and illustrated in Fig.~\ref{fig:optimization}e:
\begin{equation}
    \begin{aligned}
        P_{\max}\simeq P' \simeq s\eta_C/\tau,\quad 
        \tilde{\eta}_{\rm{MP}}\simeq \tilde{\eta}' \simeq 1-\frac{4(1-\eta_C)}{s\eta_C},
    \end{aligned}
    \label{EMP}
\end{equation}
The difference between $\tilde{\eta}_{\rm{MP}}$ and $\tilde{\eta}'$ decreases as $s$ increases, and they both converge towards the Carnot efficiency ($\tilde{\eta}=1$) for $s\to\infty$. We note that the EMP in Eq.~\eqref{EMP} recovers the result of \cite{allahverdyan2013Carnot} where the authors found the EMP by solving for the global maximum power. The power-efficiency trade-off presented in the current work provides complete performance information, such as the maximum efficiency at arbitrary power and the maximum power at arbitrary efficiency, for heat engines whose working substances exhibit degeneracy.

Equation~\eqref{EMP} indicates that the maximum power scales as $s\eta_C$ while the corresponding EMP approaches Carnot efficiency as $(s\eta_C)^{-1}$ (Figs.~\ref{fig:optimization}c and f). This scaling relation can be seen as $\lim_{s\gg 1}P_{\mathrm{max}} \simeq sk_B\Delta T/\tau$, where $\Delta T\equiv T_h-T_c$. 
For the example of interacting unit, $\sigma_i$, discussed before, $s\propto N$ in the large-$s$ regime, thus the power of each unit satisfies $P_I=P/N\sim  k_B\Delta T/\tau$.

In a recent study~\cite{shiraishi2016Universal}, Shiraishi et al. proposed a universal power-efficiency trade-off  as
\begin{equation}\label{eq:uni_bound}
    P\leq \Theta\beta_c \eta_C^2 \tilde{\eta}(1-\tilde{\eta}),
\end{equation}
where $\Theta$ relates to state and size of the heat engine. For finite-size heat engines, $\Theta$ remains finite, thus Carnot efficiency cannot be reached with finite power. However, with divergent system size, $\Theta$ may also diverge, allowing the Carnot efficiency to be approached arbitrarily closely even at maximum power, as for quantum heat engines with coherence~\cite{tajima2021Super}. In this respect, we show that also our classic engine benefits from the scaling relationships with system size. Comparing Eq.~\eqref{eq:uni_bound} with Eq.~\eqref{eq:pe_trade_expan}, we obtain the prefactor $\Theta$ as
\begin{equation}\label{eq:Thetalimit}
    \begin{cases}
        \Theta_0\equiv\lim_{\tilde{\eta}\to 0}\Theta \simeq \frac{(s-\ln s)}{\tau\eta_C\beta_c}\\
       \Theta_1 \equiv \lim_{\tilde{\eta}\to 1}\Theta \simeq \frac{\tau^{-1}s^2}{4(1-\eta_C)\beta_c}
    \end{cases}\Rightarrow\   \frac{\Theta_1}{ \Theta_0} \sim s \;,
\end{equation}
with the ratio of the two prefactors scaling linearly with $s$ in the large-$s$ limit. Note that, although we have shown the connection between the trade-off we obtained and the universal power-efficiency trade-off~\cite{shiraishi2016Universal}, $\Theta_0\neq\Theta_1$ implies that, for heat engines with divergent physical quantities near the phase transition point, it is not possible to characterize the performance of the engine over the entire range of $\tilde{\eta}$ using a unified $\Theta$. In other words, $\Theta$ is an implicit function of $\tilde{\eta}$~\cite{tight-analytical_tradeoff}.

As a final remark, the presented results clearly demonstrate that the performance of our biochemical engine exceeds the universal power-efficiency trade-off obtained in the low-dissipation regime~\cite{holubec2016Maximum,ma2018Universal,zhai2023experimental}. In this regime, the trade-off relation is determined by the typical irreversible entropy generation $S^{(\rm ir)} = \Sigma/t$, where $t$ is proportional to cycle duration~\cite{esposito2010Efficiency,martinez2016brownian,ma2020experimental}. For non-interacting systems, both $\Sigma$ and the reversible entropy change, $\Delta S$, increase with the system size~\cite{ma2018Universal,ma2020experimental,yuan2022Optimizing}. Consequently, the only way to reduce energy dissipation is to slow down the cycle operation, resulting in higher efficiency but lower output power. However, for many-body interacting systems, the relative value of $\Sigma$ compared to $\Delta S$ can effectively be reduced by increasing the system size~\cite{abiuso2020Optimal,rolandi2023Collective}, enabling high-speed cycles that achieve Carnot efficiency when $S^{\rm{(ir)}}/\Delta S\rightarrow 0$. This observation is consistent with our results: in the geometric representation of the cycle (Figs.~\ref{fig:optimization}a and d),  $\bar{W}_\mathrm{slow}$ and $\bar{W}_\mathrm{fast}$ respectively correspond to $S^{(\rm ir)}$ and $\Delta S$. For a given cycle duration, $\bar{W}_\mathrm{slow}/\bar{W}_\mathrm{fast}$ decreases with $s$, eventually vanishing when $s\to\infty$. 

\textit{Breaking the 1/2-universality of EMP}.---
For a heat engine operating in the linear irreversible regime ($\Delta T/T_h\ll 1$), the leading-term coefficient of EMP with respect to $\eta_C$ is $\alpha=1/2$, irrespectively of engine details~\cite{vandenbroeck2005Thermodynamic,tu2008Efficiency,yuan2022Optimizing,zhai2023experimental}. However, this universality~\cite{Tu2012Recent} relies on the validity of linear-irreversible thermodynamics and can be violated in the presence of divergent quantities \cite{seifert2011Efficiency,lee2018Nonuniversality}. Indeed, our minimal heat engine shows a violation of the $1/2$-universality when $s\to \infty$ due to the divergent slope of the equilibrium population curves, as demonstrated by Eq.~\eqref{eq:response}. We notice that the order of taking $s\to\infty$ and $\eta_C\to\ 0$ significantly affects the result:
\begin{equation}
    \begin{cases}
    \lim_{\eta_C\to 0}\lim_{s\to \infty}\eta_\mathrm{MP}=\eta_C\\
    \lim_{s\to \infty}\lim_{\eta_C\to 0}\eta_\mathrm{MP}=\eta_C/2.
    \end{cases}
\end{equation}
This implies that, in the parametric space $(\eta_C,s)$, different limiting paths lead to different EMPs. By varying the limiting path, $\alpha$ transitions continuously from 1/2 to 1, as shown by the colormap of $\tilde{\eta}_\mathrm{MP}\equiv\eta_\mathrm{MP}/\eta_C$ in Fig.~\ref{fig:1/2-universality}a. 

For finite-size working substances, one can always choose a small enough $\eta_C$ to have the corresponding EMP approaching $\eta_C/2$ (Fig.~\ref{fig:1/2-universality}b) \cite{supplemental_material}, thus the universality of $\lim_{\eta_C\rightarrow0}\eta_\mathrm{MP}=\eta_C/2$ holds in the finite-size case. Consequently, the criterion to determine whether a heat engine is operating in the linear irreversible regime should be $s\eta_C\ll1$ instead of the conventional $\eta_C\ll1$.

\textit{Conclusion}.---
In this work, we propose a minimal model that highlights the absence of a general power-efficiency trade-off relation for heat engines with divergent quantities. The discrepancy between the EMP and Carnot efficiency is dictated by the system size, which is characterized by the number of microscopic states. Remarkably, Carnot efficiency at maximum power is attained in the thermodynamic limit with infinite system size. This observation aligns with recent findings on phase-transition-enhanced performance of heat engines \cite{campisi2016Power,vroylandt2018Collective}. 

Our minimal model also reveals the beneficial impact on heat engine performance of engineering microscopic states to form two coarse-grained states with distinct degeneracies, reinforcing earlier findings on the role of degeneracy and entropy in thermal systems \cite{golubeva2013Entropygenerated,esposito2010Extracting,correa2015Individual,ma2017quantum,tajima2021Super}. These two states can be regarded as ordered and disordered states during a phase transition, achieved through collective effects in many-body interacting systems \cite{campisi2016Power,ma2017quantum,herpich2018Collective,rolandi2023Collective,vroylandt2018Collective,filho2023Powerful}. By focusing on the resulting energy-level structure rather than delving into the intricate interactions within many-body systems, our model succinctly and intuitively demonstrates how collective effects enhance the performance of heat engines. 
By introducing the idea of designing optimal internal structure for the engines' working substances, our approach will open up new avenues for optimizing heat engines, going beyond the conventional pursuit of optimal cycle control scheme \cite{ma2018optimal,yuan2022Optimizing,frim2022geometric,chen2022microscopic}.


\begin{acknowledgments}
    \textit{Acknowledgments}.--- The authors thank Simone Pigolotti for comments on a preliminary version of this manuscript. S.L. thanks the JSPS Strategic Fellowship for support under grant No. GR22106. S.L. and P.D.L.R. thank the Swiss National Science Foundation for support under grant 200020\_178763. Y.H.M. thanks the National Natural Science Foundation of China for support under grant No. 12305037 and the Fundamental Research Funds for the Central Universities for support under grant No. 2023NTST017.
\end{acknowledgments}

\bibliography{refs}

\begin{thebibliography}{55}
\bibitem [{supplemental_material({\natexlab{b}})}]{optimization}%
  \BibitemOpen
  \bibinfo {note} {At fixed temperatures and $s$, we numerically optimize power
  with respect to $\epsilon_h$ and $\epsilon_c$ to find the efficiency at
  maximum power. For the power-efficiency trade-off, we substitute $\epsilon_c
  = \epsilon_h(1-\eta_C\tilde{\eta})$ and optimize power over $\epsilon_h$
  under different $\tilde{\eta}$.}\BibitemShut {Stop}%
\bibitem [{tight-analytical_tradeoff()}]{tight-analytical_tradeoff}%
  \BibitemOpen
  \bibinfo {note} {This finding leaves an open question: Does a tight
  analytical tradeoff between power and efficiency exist throughout the entire
  range of $\tilde{\eta}$? We have attempted to find it but have not been
  successful so far. The asymptotic expressions in
  Eq.~\eqref{eq:pe_trade_expan} at both ends of $\tilde{\eta}$ can serve as an
  a good approximation for the complete tradeoff relation in the large-$s$
  regime.}\BibitemShut {Stop}%
\end{thebibliography}

\end{document}


\begin{CJK*}{UTF8}{gbsn}

\title{Supplemental Materials for: \\ ``A Minimal Model for Carnot Efficiency at Maximum Power''}
\author{Shiling Liang (梁师翎)}
 \email{shiling.liang@epfl.ch}
\affiliation{Institute of Physics, School of Basic Sciences, \'Ecole Polytechnique F\'ed\'erale de Lausanne (EPFL), 1015 Lausanne, Switzerland}
\affiliation{Biological Complexity Unit, Okinawa Institute of Science and Technology Graduate University, Onna, Okinawa 904-0495, Japan}
\author{Yu-Han Ma}
 \email{yhma@bnu.edu.cn}
\affiliation{School of Physics and Astronomy, Beijing Normal University, Beijing, 100875, China}
\affiliation{Graduate School of China Academy of Engineering Physics, No. 10 Xibeiwang East Road, Haidian District, Beijing, 100193, China}
\author{Daniel Maria Busiello}
\affiliation{Max Planck Institute for the Physics of Complex Systems, 01187 Dresden, Germany}
\author{Paolo De Los Rios}
\affiliation{Institute of Physics, School of Basic Sciences, \'Ecole Polytechnique F\'ed\'erale de Lausanne (EPFL), 1015 Lausanne, Switzerland}
\affiliation{Institute of Bioengineering, School of Life Sciences, \'Ecole Polytechnique F\'ed\'erale de Lausanne (EPFL), 1015 Lausanne, Switzerland}

\maketitle
\end{CJK*}
\onecolumngrid
This document provides the detailed derivations and the supporting discussions to the main text of the Letter. The contents of the Supplemental Materials are listed as follows

\tableofcontents

\section{\label{Stadystate}The stationary state}
The master equation describing the biochemical engine is
\begin{equation}\label{eq:master_SI}
    \begin{aligned}
       \frac{d}{dt}p_{u} &= (k_{h-} + k_{c-})p_{v}- (k_{h+}+ k_{c+})p_u ,\\
       \frac{d}{dt}p_{v} &=  (k_{h+}+ k_{c+})p_u-(k_{h-} + k_{c-})p_{v} \;.
    \end{aligned}
\end{equation}
Its stationary distribution satisfies
\begin{equation}
p_{v}^\mathrm{ss}=\frac{k_{h+}+k_{c+}}{k_{h-}+k_{c-}}p_u^\mathrm{ss},
\end{equation}
with the normalization condition $p_u^\mathrm{ss}+p_{v}^\mathrm{ss} = 1$. One can find the stationary distribution of the probability to be in the state $U$ as
\begin{equation}
    \begin{aligned}
        p^\mathrm{ss}_{u}
        =\frac{k_{h-}+k_{c-}}{k_{h+}+k_{h-}+k_{c+}+k_{c-}}
        =\frac{k_h}{k_c+k_h}\underbrace{\frac{k_{h-}}{k_{h}}}_{\pi_u^h}+\frac{k_c}{k_c+k_h}\underbrace{\frac{k_{c-}}{k_{c}}}_{\pi_u^c},
    \end{aligned}
\end{equation}
where we define the characteristic rate in the two reservoirs as $k_h = k_{h-}+k_{h+}$ and $k_c = k_{c-}+k_{c+}$. The stationary value of $p_{v}^\mathrm{ss}$ can be obtained in the same fashion. The steady-state flux is 
\begin{equation}
    \begin{aligned}
        J^\mathrm{ss}
        &=k_{h+} p_u^\mathrm{ss}-k_{h-}p_{v}^\mathrm{ss}\\
        &= \frac{k_{h+}(k_{h-}+k_{c-})-k_{h-}(k_{h+}+k_{c+})}{k_c+k_h}\\
        &= \frac{k_ck_h}{(k_c+k_h)}\left(\frac{k_{h+}k_{c-}}{k_ck_h}-\frac{k_{h-}k_{c+}}{k_ck_h}\right)\\
        &= \frac{k_ck_h}{(k_c+k_h)}\left(\frac{k_{h+}}{k_h}\left(1-\frac{k_{c+}}{k_c}\right)-\left(1-\frac{k_{h+}}{k_h}\right)\frac{k_{c+}}{k_c}\right)\\
        &= \frac{k_ck_h}{(k_c+k_h)}\left(\frac{k_{h+}}{k_h}-\frac{k_{c-}}{k_c}\right)\\
        &= \frac{1}{\tau_h+\tau_c}(\pi_{v}^{h}-\pi_{v}^c).
    \end{aligned}
\end{equation}
where we introduce the two timescales of transitions in the hot and cold reservoir as $\tau_h =k_h^{-1}  $ and $\tau_c=k_c^{-1}$.

\section{\label{Otto cycle}Comparison between the Otto Cycle and Steady-State Heat Engine}

\subsection{The timescale of Otto cycle.} 
\begin{figure}[!htb]
\includegraphics[width=.5\columnwidth]{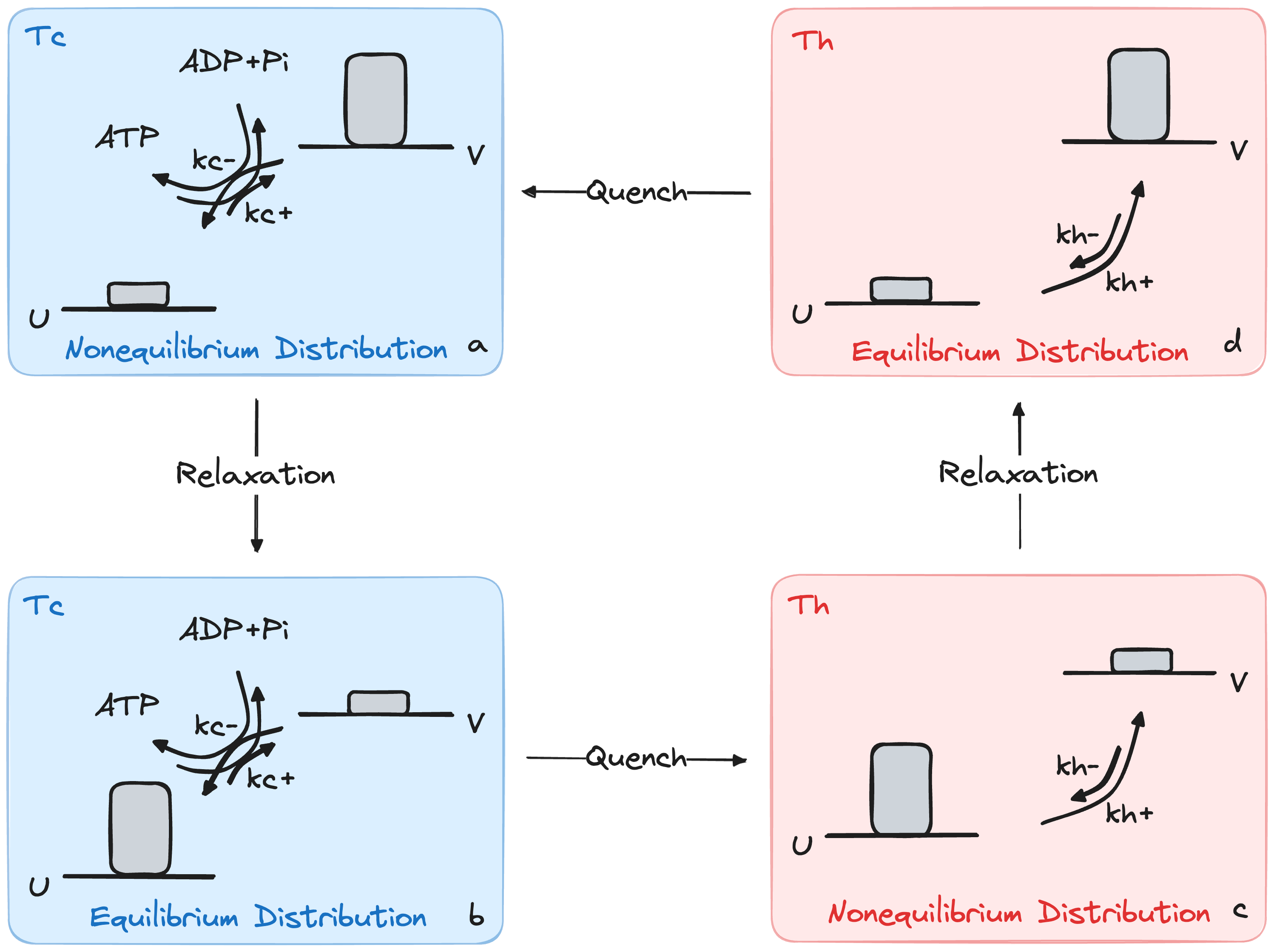}
\caption{Illustration of the Otto cycle as discussed in the main text. The diagram shows the probability distribution between two states, represented by the bars on states $U$ and $V$. }
    \label{fig:Otto_cycle}
\end{figure}

The Otto cycle discussed in the main text can be decomposed into four processes: two rapid changes (quenches), assumed to be instantaneous, and two relaxations, as illustrated in Fig.~\ref{fig:Otto_cycle}. The two relaxation processes are governed by transition rates and, therefore, are characterized by a typical timescale. In particular, for the relaxation from state $c$ to state $d$, we have 
\begin{equation}
\begin{aligned}
    \frac{dp_v}{dt} = -k_{h-} p_v+k_{h+} p_u,\quad 
    \frac{dp_u}{dt} = k_{h-} p_v-k_{h+} p_u.
    \end{aligned}
\end{equation}
The solution of such a system reads 
\begin{equation}
    p_v(t) = \pi_v^{h}(1-e^{-(k_{h+}+k_{h-})t}),\quad p_u(t) = \pi_u^{h}(1-e^{-(k_{h+}+k_{h-})t})
\end{equation}
where $\pi_v^{h} = k_{h+}/(k_{h-}+k_{h+})$ and $\pi_u^{h} = k_{h-}/(k_{h-}+k_{h+})$ are the equilibrium distributions. From this solution, we find that the relaxation timescale is $\tau_h = 1/(k_{h+}+k_{h-})$, as suggested in the main text. For the relaxation from $b$ to $c$, we can follow the same procedure to find the other relaxation time scale $\tau_c = 1/(k_{c+}+k_{c-})$. Combining these two relaxation timescales gives the period of Otto cycle as $\tau =\tau_h+\tau_c$.

\subsection{The power of the Otto cycle.}
The output work of the Otto cycle can be computed from the two quenching processes. For the quench $d\to a$, ATP-hydrolysis is coupled to the transition between states, leading to the energy gap change from $\epsilon_h = \epsilon_V-\epsilon_U$ to $\epsilon_c = \epsilon_V-\epsilon_U-\Delta\mu$, where $\epsilon_U$ and $\epsilon_V$ are the energies of the states $U$ and $V$, respectively. This quenching process is associated with the output work 
\begin{equation}
    W_{d\to a} = \pi_v^{h}(\epsilon_h-\epsilon_c).
\end{equation}
While for the quench from $b$ to $c$, the output work is 
\begin{equation}
    W_{b\to c} =  -\pi_v^{c}(\epsilon_h-\epsilon_c).
\end{equation}
By combining them, we obtain the total output work per Otto cycle
\begin{equation}
    W_{\rm Otto} = W_{b\to c}+ W_{d\to a} = (\pi_v^h-\pi_v^c)(\epsilon_h-\epsilon_c).
\end{equation}
Remember that the period of the Otto cycle is $\tau = \tau_h+\tau_c$. By dividing the output work by the cycle period, we get the output power as 
\begin{equation}
    P_{\rm Otto} = \tau^{-1} (\pi_v^h-\pi_v^c)(\epsilon_h-\epsilon_c),
\end{equation}
which has the same form of the steady-state output power in the bipartite system given in the main text.

An alternative method to calculate the output work involves calculating the amount of synthesized ATP. During the relaxation process $a\to b$, the change in the amount of ATP can be expressed as $\Delta p_v= \pi_v^{h}-\pi_v^c$, and the free energy change from ADP+Pi to ATP is $\Delta\mu = \epsilon_h-\epsilon_c$. We thus have the output work in terms of ATP synthesis
\begin{equation}
    W_{\rm {ADP+Pi\to ATP}} =\Delta\mu\Delta p_v = (\pi_v^{h}-\pi_v^c)(\epsilon_h-\epsilon_c) = W_{\rm Otto}.
\end{equation}
which is the same as the output work computed from the two quenching processes. 

\subsection{\label{Appendix Qof Otto cycle}The quality of the Otto cycle.}
The extractable work after the two Otto quenches are the remained amount of information encoded in the non-equilibrium distributions, which can be computed as the non-equilibrium free energy difference
\begin{equation}
\Delta \mathcal{F}_{cd} =  \epsilon_h (\pi_v^c-\pi_v^h)-T_h(S_c-S_h)=-\epsilon_h\Delta\pi+T_h\Delta S,
\end{equation}
\begin{equation}
\Delta \mathcal{F}_{ab} = \epsilon_c(\pi_v^h-\pi_v^c)-T_c(S_h-S_c)=\epsilon_c\Delta\pi-T_c\Delta S.
\end{equation}
where $\Delta\pi = \pi_v^h-\pi_v^c$, and $\Delta S = S_h-S_c$. When we calculate the entropy change, we need to be aware that the system is coarse-grained so that part of the system entropy is hidden in the degeneracy. For a given coarse-grained 2-state distribution $(p_u,p_v)$, the exact entropy is 
\begin{equation}
S(p_u,p_v) = -p_u\ln p_u - p_v\ln p_v + \underbrace{p_u \ln n_u+p_v \ln n _v}_{p_v s+\ln n_u}.
\end{equation}
We denote the first part as $S_0=-p_u\ln p_u - p_v\ln p_v$. Therefore we can find that the slow-work is
\begin{equation}
W_\mathrm{slow}=\Delta \mathcal{F}_{cd}+\Delta \mathcal{F}_{ab}=-\Delta\epsilon\Delta\pi+\Delta T\Delta S_0+\Delta\pi s\Delta T,
\end{equation}
and the fast-work is
\begin{equation}
W_\mathrm{fast}=(\epsilon_h-\epsilon_c)(\pi_v^h-\pi_v^c)=\Delta\epsilon\Delta\pi,
\end{equation}
where $\Delta\epsilon = \epsilon_h-\epsilon_c$, $\Delta T=T_h-T_c$, and $\Delta S_0 = S_0^h-S_0^c$. The quality which quantifies the fraction of fast-work is:
\begin{equation}
 \mathcal{Q}\equiv \frac{W_\mathrm{fast}}{W_\mathrm{fast}+W_\mathrm{slow}} = \frac{\Delta\epsilon\Delta\pi}{\Delta\pi s\Delta T+\Delta T\Delta S_0}.
\end{equation}
When $s$ is large, the coarse-grained states contribute the most to the entropic change thus we can drop the $\Delta S_0$ term. Thus, the quality factor is given by
\begin{equation}
\lim_{s\gg 1} \mathcal{Q} = \frac{\Delta \epsilon }{s\Delta T}=\frac{\bar{\epsilon}_h-\bar{\epsilon}_c}{T_h-T_c}.
\end{equation}

\section{\label{Appendix tradeoff}Power-efficiency trade-off relation}


The general power-efficiency trade-off with different $s$ is numerically plotted in Fig.~2f of the main text. It is clearly seen in this figure that for an arbitrary given power, the maximum achievable efficiency increases with $s$. Hence, we reasonably speculate that the upper branch of the trade-off will become a plateau ($\eta = \eta_C$) at the thermodynamic limit of $s\rightarrow\infty$. This means that maximum power and maximum efficiency can be achieved simultaneously. To analytically prove this result, we focus on this trade-off relation under two limits of $\eta$. Here we set the normalized efficiency $\tilde{\eta}\equiv \eta/\eta_C$ and $\beta_h=1$, so that we can rewrite $\epsilon_c = \epsilon_h(1-\eta)= \epsilon_h(1-\tilde{\eta}\eta_C)$ and $\beta_c = (1-\eta_C)^{-1}$. Substituting these expressions into the Eq.(4) in the main text, we have
\begin{equation}\label{eq:Pofeta_SI}
\begin{aligned}
   P 
   &=J^{\rm ss} (\epsilon_h-\epsilon_c)\\
   &=\tau^{-1} (\epsilon_h-\epsilon_c)(\pi_v^h-\pi_v^c) &\hookleftarrow^{\text{Eq.(4) in main text}}\\
   &= \tau^{-1}(\epsilon_h-\epsilon_c)\left(\frac{1}{e^{\beta_h\epsilon_h - s}+1}-\frac{1}{e^{\beta_c\epsilon_c - s}+1}\right)&\hookleftarrow^{\text{Using the explicit forms of $\pi_v^h$ and $\pi_v^c$}}\\ 
    &= \tau^{-1}\epsilon_h\tilde{\eta}\eta_C\left(\frac{1}{e^{\epsilon_h-s}+1}-\frac{1}{e^{{\frac{\epsilon_h\left(1-\tilde{\eta}\eta_C\right)}{1-\eta_C}-s}}+1}\right)&\hookleftarrow^{\text{Change of variables}}
\end{aligned}
\end{equation}

\paragraph{$\tilde{\eta}\to 1$ limit}
In this limit, we expand the power in Eq.~\eqref{eq:Pofeta_SI} in terms of $\sigma\equiv (1-\tilde{\eta})$ up to the first order as
\begin{equation}\label{eq:P_expan_1}
\begin{aligned}
   P 
   &=\tau^{-1}\epsilon_h\eta_C(1-\sigma)\left(\frac{1}{e^{\epsilon_h-s}+1}-\frac{1}{e^{\frac{\epsilon_h(1-\eta_C(1-\sigma))}{1-\eta_C}-s}+1}\right)\\
   &=\tau^{-1}\epsilon_h\eta_C(1-\sigma)\left(\frac{1}{e^{\epsilon_h-s}+1}-\frac{1}{e^{\epsilon_h-s}e^{\frac{\epsilon_h\eta_C\sigma}{1-\eta_C}}+1}\right)\\
   &=\tau^{-1}\epsilon_h\eta_C(1-\sigma)\left(\frac{1}{e^{\epsilon_h-s}+1}-\frac{1}{e^{\epsilon_h-s}(1+\frac{\epsilon_h\eta_C\sigma}{1-\eta_C})+1+\mathcal{O}(\sigma^2)}\right)\\
   &=\tau^{-1}\epsilon_h\eta_C(1-\sigma)\left(\frac{1}{e^{\epsilon_h-s}+1}-\frac{1}{e^{\epsilon_h-s}+1}+\frac{e^{\epsilon_h-s}\epsilon_h\eta_C}{(e^{\epsilon_h-s}+1)^2(1-\eta_C)}\sigma+\mathcal{O}(\sigma^2)\right)\\
   &=\frac{\tau^{-1}\eta_C^2 \varepsilon_h^2 e^{\varepsilon_h-{s} }}{\left(1-\eta_C\right) \left(e^{\varepsilon_h-{s} }+1\right)^2}\sigma+\mathcal{O}(\sigma^2),
   \end{aligned}
\end{equation}
where we use the high-temperature rescaled energy level $\varepsilon_h\equiv \epsilon_h/T_h$ to simplify the expression. The trade-off curve is obtained by computing the maximized linear coefficient. For fixed entropy difference, we optimize the coefficient with respect to $\varepsilon_h$, namely
\begin{equation}
\frac{d}{d\varepsilon_h}\left(\frac{\tau^{-1}\eta_C^2 \varepsilon_h^2 e^{\varepsilon_h-{s} }}{\left(1-\eta_C\right) \left(e^{\varepsilon_h-{s} }+1\right)^2}\right)=0.
\end{equation}
The above equality leads to 
\begin{equation}\label{eh}
    e^{\varepsilon_h-{s}} =\frac{\varepsilon_h+2}{\varepsilon_h-2}.
\end{equation}
In the large degeneracy regime, one has $\varepsilon_h>{s}\gg 1$, such that $(\varepsilon_h+2)/(\varepsilon_h-2)\simeq 1$. Therefore, the solution of Eq.~\eqref{eh} is approximately $\varepsilon_h = {s}$ ($e^{\varepsilon_h-{s}}\simeq 1$). Substituting it back into Eq.~\eqref{eq:P_expan_1} yields
\begin{equation}\label{eq:P_expan_1_mp}
    \lim_{\tilde{\eta}\to 1}P \simeq \frac{\tau^{-1}{s} ^2 \eta_C^2}{4(1- \eta_C)} (1-\tilde{\eta}) 
\end{equation}
 
\paragraph{$\tilde{\eta}\to 0$ limit} 
In this limit, we can expand the power in Eq.~\eqref{eq:Pofeta_SI} with respect to $\tilde{\eta}$ to get the first order as 
\begin{equation}\label{eq:P_expan_0}
   \begin{aligned}
   P 
   &=\tau^{-1}\eta_C\epsilon_h\tilde{\eta}\left(\frac{1}{e^{\epsilon_h-s}+1}-\frac{1}{e^{{\frac{\epsilon_h}{1-\eta_C}-s}}e^{-{\frac{\epsilon_h\left(\tilde{\eta}\eta_C\right)}{1-\eta_C}}}+1}\right)\\
   &=\tau^{-1}\eta_C\epsilon_h\tilde{\eta}\left(\frac{1}{e^{\epsilon_h-s}+1}-\frac{1}{e^{{\frac{\epsilon_h}{1-\eta_C}-s}}(1+\mathcal{O}(\tilde{\eta}))+1}\right)\\
   &= \tau^{-1}\eta_C \varepsilon _h \left(\frac{1}{e^{\varepsilon _h-{s} }+1}-\frac{1}{e^{\frac{\varepsilon _h}{1-\eta_C}-{s} }+1}\right)\tilde{\eta} + \mathcal{O}(\tilde{\eta}^2).
   \end{aligned}
\end{equation}
With the change of variable $x\equiv\varepsilon_{h}-{s}$, Eq.~\eqref{eq:P_expan_0} is re-written
as
\begin{equation}
P\simeq \tau^{-1}\eta_C\left({s}+x\right)\left(\frac{1}{e^{x}+1}-\frac{1}{e^{\frac{x+\eta_C{s}}{1-\eta_C}}+1}\right)\tilde{\eta}.
\end{equation}
Optimizing $P$ with respect to $x$, namely, 
\begin{equation}
\frac{dP}{dx}=\tau^{-1}\eta_C\tilde{\eta}\frac{d}{dx}\left[\left({s}+x\right)\left(\frac{1}{e^{x}+1}-\frac{1}{e^{\frac{x+\eta_C{s}}{1-\eta_C}}+1}\right)\right]=0,
\end{equation}
leads to 
\begin{equation}
\frac{1}{e^{x}+1}-\frac{1}{e^{\frac{x+\eta_C{s}}{1-\eta_C}}+1}+\left({s}+x\right)\left[\frac{-e^{x}}{\left(e^{x}+1\right)^{2}}-\frac{-e^{\frac{x+\eta_C{s}}{1-\eta_C}}}{\left(1-\eta_C\right)\left(e^{\frac{x+\eta_C{s}}{1-\eta_C}}+1\right)^{2}}\right]=0.\label{eq:x*solution}
\end{equation}
We also focus on the large degeneracy regime (${s}\gg1$) in which we intuitively guess that the solution of the above equality, i.e. $x=x^{*}$, satisfies 
\begin{equation}
1\ll\left|x^{*}\right|\ll{s},\label{eq:assumex*}
\end{equation}
such that we have
\begin{equation}
\frac{1}{e^{\frac{x+\eta_C{s}}{1-\eta_C}}+1}\approx0,\frac{e^{\frac{x+\eta_C{s}}{1-\eta_C}}}{\left(1-\eta_C\right)\left(e^{\frac{x+\eta_C{s}}{1-\eta_C}}+1\right)^{2}}\approx0,{s}+x\approx{s}.
\end{equation}
Then, Eq.~\eqref{eq:x*solution} is approximated as
\begin{equation}
\frac{1}{e^{x}+1}-\frac{{s} e^{x}}{\left(e^{x}+1\right)^{2}}=0.
\end{equation}
The solution of the above equality is $e^{x}=1/({s}-1)$, namely
\begin{equation}
x^{*}=-\ln\left({s}-1\right)\simeq-\ln{s}.
\end{equation}
This solution is consistent with the assumption in Eq.~\eqref{eq:assumex*}.
Finally, we obtain the optimal $\varepsilon_{h}$ in the small $\tilde{\eta}$
and large ${s}$ regimes as
\begin{equation}
\varepsilon_{h}=\varepsilon_{h}^{*}={s}-\ln{s}.
\end{equation}
Substituting it into Eq.~\eqref{eq:P_expan_0} and taking the large-${s}$ limit to further simplify the coefficient, we obtain
\begin{equation}\label{eq:P_expan_0_mp}
  \begin{aligned}
  \lim_{\tilde{\eta}\to 0}P
     &= \tau^{-1}(s-\ln s)\eta_C\tilde{\eta}\Bigg(\frac{1}{e^{\ln s}+1}-\overbrace{\frac{1}{e^{\frac{s-\ln s}{1-\eta_C}-s}+1}}^{\text{goes to $0$ in the large-$s$ limit}}\Bigg)\\
     &\simeq\tau^{-1}\frac{{s}  ({s} -\ln{s} )\eta_C }{{s} +1} \tilde{\eta}.
     \end{aligned}
\end{equation}
These two linear expansions (Eq.~\eqref{eq:P_expan_1_mp} and Eq.~\eqref{eq:P_expan_0_mp}) give an upper bound of the power-efficiency trade-off curve. Therefore, the upper bounds of the rescaled efficiency at maximum power and of the maximum power can be estimated by the intersection of these two lines, denoted as $(\tilde{\eta}', P')$, which is simply obtained as follows: 
\begin{equation}
    \tilde{\eta}'\simeq \frac{({s} +1) \eta_C}{({s} -3) \eta_C+4}\simeq 1- \frac{4 \left(1-\eta_C\right)}{  \eta_C}{s}^{-1},
\end{equation}
and $P'\simeq\tau^{-1}\eta_C{s}$, respectively. Note that the maximum power increases linearly with ${s}$ while the EMP approaches to Carnot efficiency being inversely proportional to ${s}$. This means that increasing the degeneracy of the system can improve power and maximum power efficiency of the engine at the same time, which is a perfect advantage. 

Moreover, numerical evidences suggest that, besides $\tilde{\eta}=1$, the linear expansion in Eq.~\eqref{eq:P_expan_0_mp} can be a good approximation of the power-efficiency tradeoff in the interval $0\leq\tilde{\eta}<1$. In this sense, the normalized approximate trade-off becomes
\begin{equation}\label{eq:P_expan_0_mp_2}
     \tilde{P}\equiv \frac{P}{P'}\leq \left(1-\frac{\ln{s}}{{s}}\right) \tilde{\eta}.
\end{equation}
Geometrically, the approximate tradeoff relation becomes an isosceles right triangle in the normalized power-efficiency parametric space since $\lim_{{s}\rightarrow\infty}d \tilde{P}/d\tilde{\eta}=1$.
This is intuitively reflected in the tradeoff curve associated with ${s}=10^4$ in Fig.~2f of the main text.

\section{\label{Appendix EMPlimit}EMP in the limits $\eta_C\rightarrow0$ and $\eta_C{s}\rightarrow0$}

We rewrite the third line of Eq.~\eqref{eq:Pofeta_SI} using the change of variable $\theta\equiv \epsilon_h\beta_c$ ,
\begin{equation}
\begin{aligned}
P
&=\tau^{-1}\epsilon_h (1-\frac{\epsilon_c}{\epsilon_h})\left(\frac{1}{e^{\epsilon_h\epsilon_c\frac{\beta_h}{\beta_c}-s}+1}-\frac{1}{e^{\epsilon_h\beta_c\frac{\epsilon_c}{\epsilon_h}-s}+1}\right) \\
&=\tau^{-1}\beta_c^{-1}\theta\left(1-\frac{\epsilon_c}{\epsilon_h}\right) \left(\frac{1}{e^{\theta\frac{\beta_h}{\beta_c}-s}+1}-\frac{1}{e^{\theta\frac{\epsilon_c}{\epsilon_h}-s}+1}\right) &\hookleftarrow{\text{let $\theta \equiv\epsilon_h\beta_c$}}\\ 
&=\tau^{-1}\beta_c^{-1}\theta\eta \left(\frac{1}{e^{\theta(1-\eta)-s}+1}-\frac{1}{e^{\theta(1-\eta_C)-s}+1}\right) &\hookleftarrow{\text{using $\eta=1-\frac{\epsilon_c}{\epsilon_h}$ and $\eta_C=1-\frac{\beta_h}{\beta_c}$ }}
\end{aligned}
\end{equation}
In the limits of $\eta_C\rightarrow0$ and $\eta_C{s}\rightarrow0$,
it is natural to assume that $\eta\theta\leq\eta_C\theta\ll1$,
such that
\begin{align}
P=&\tau^{-1}\beta_c^{-1}\eta\theta\left[\frac{1}{1+e^{\theta-{s}}(1-\eta_C\theta)}-\frac{1}{1+e^{\theta-{s}}(1-\eta\theta)}\right]\\
 & \simeq\frac{\tau^{-1}\beta_c^{-1}\eta\theta\left[e^{\theta-{s}}(1-\eta\theta)-e^{\theta-{s}}(1-\eta_C\theta)\right]}{\left[1+e^{\theta-{s}}(1-\eta_C\theta)\right]\left[1+e^{\theta-{s}}(1-\eta\theta)\right]}\\
 & \simeq\frac{\tau^{-1}\beta_c^{-1}e^{\theta-{s}}\theta^{2}\eta\left(\eta_C-\eta\right)}{\left(1+e^{\theta-{s}}\right)^{2}}.
\end{align}
By optimizing $P$ with respect to $\eta$ directly gives
\begin{equation}
\frac{\partial P}{\partial\eta}=0\rightarrow\eta_C-2\eta=0,
\end{equation}
which indicates that the efficiency at maximum power is
\begin{equation}
\lim_{\eta_C\rightarrow0}\eta_{\mathrm{MP}}=\frac{\eta_C}{2}
\end{equation}